\documentclass[reprint, amsmath, amssymb, aps, prb, longbibliography, superscriptaddress, nofootinbib, floatfix]{revtex4-2}
\usepackage[american]{babel}
\usepackage[utf8]{inputenc}
\usepackage{csquotes}
\MakeOuterQuote{"}
\usepackage{dcolumn}
\usepackage{bm}
\usepackage{placeins}
\usepackage{graphicx}
\usepackage[caption=false]{subfig}
\usepackage{braket}
\usepackage[colorlinks, bookmarks=false, citecolor=blue, linkcolor=red, urlcolor=blue]{hyperref}
\usepackage{times}

\DeclareMathOperator{\sign}{sign}

\DeclareSymbolFont{rmlargesymbols}{U}{euex}{m}{n}
\DeclareMathSymbol{\rmintop}{\mathop}{rmlargesymbols}{82}
\newcommand{\rmint}{\rmintop\nolimits}

\newcommand{\eup}{\mathrm{e}}
\newcommand{\iup}{\mathrm{i}}
\newcommand{\dup}{\mathrm{d}}

\newcommand{\Heff}[0]{H_\mathrm{Fl}}

\begin{document}
\title{Estimating heating times in periodically driven quantum many-body systems via avoided crossing spectroscopy}

\author{Artem Rakcheev}
\email{artem.rakcheev@psi.ch}
\affiliation{Institut für Theoretische Physik, Universität Innsbruck, A-6020 Innsbruck, Austria}
\affiliation{Laboratory for Theoretical and Computational Physics, Paul Scherrer Institut, 5232 Villigen PSI, Switzerland}

\author{Andreas M. Läuchli}
\affiliation{Institut für Theoretische Physik, Universität Innsbruck, A-6020 Innsbruck, Austria}
\affiliation{Laboratory for Theoretical and Computational Physics, Paul Scherrer Institut, 5232 Villigen PSI, Switzerland}
\affiliation{Institute of Physics, Ecole Polytechnique Fédérale de Lausanne (EPFL), CH-1015 Lausanne, Switzerland}

\date{\today}
\begin{abstract}
Periodic driving of a quantum (or classical) many-body system can alter the systems properties significantly and therefore has emerged as a promising way to engineer exotic quantum phases, such as topological insulators and discrete time crystals. A major limitation in such setups, is that generally interacting, driven systems will heat up over time and lose the desired properties. Understanding the relevant time scales is thus an important topic in the field and so far, there have only been few approaches to determine heating times for a concrete system quantitatively, and in a computationally efficient way. In this article we propose a new approach, based on building the heating rate from microscopic processes, encoded in avoided level crossings of the Floquet propagator. We develop a method able to resolve individual crossings and show how to construct the heating rate based on these. The method is closely related to the Fermi Golden Rule approach for weak drives, but can go beyond it, since it captures non-perturbative effects by construction. This enables our method to be applicable in scenarios such as the heating time of discrete time crystals or frequency dependent couplings, which are very relevant for Floquet engineering, where previously no efficient methods for estimating heating times were available.
\end{abstract}

\maketitle
\section{\label{sec:introduction}Introduction}
Periodically driven classical and quantum systems (also called Floquet systems, after the French mathematician Gaston Floquet, in this context) have been studied since the birth of those theories. In recent decades the study of periodically driven quantum many-body systems (QMBS) (for which we shall simply use the term Floquet systems hereafter) gathered interest, due to theoretical and experimental developments \cite{eckardt_colloquium_2017, oka_floquet_2019} including drive-assisted tunneling~\cite{grifoni_driven_1998} leading to the observation of dynamical localization in Bose-Einstein condensates in shaken optical lattices~\cite{eckardt_superfluidinsulator_2005, lignier_dynamical_2007, eckardt_exploring_2009}, the photovoltaic Hall effect 
in graphene \cite{oka_photovoltaic_2009, kitagawa_transport_2011}, the realization of topological models by engineering spin-orbit couplings and artificial gauge fields~\cite{kitagawa_topological_2010, aidelsburger_experimental_2011, kolovsky_creating_2011, aidelsburger_realization_2013, miyake_realizing_2013, struck_engineering_2013, kennedy_spinorbit_2013, jotzu_experimental_2014, goldman_lightinduced_2014, aidelsburger_measuring_2015, bermudez_interactiondependent_2015, harper_topology_2020}, the simulation of lattice gauge theories~ \cite{schweizer_floquet_2019} and the observation of discrete time crystals~\cite{choi_observation_2017, zhang_observation_2017, yao_time_2018, else_discrete_2020}.  Further potential prospects include the realization of Hopf insulators~\cite{park_higherorder_2019, schuster_floquet_2019, unal_hopf_2019}, Floquet engineering using trapped ions~\cite{bermudez_photonassistedtunneling_2012, grass_chiral_2018, kiefer_floquetengineered_2019}, counter-diabatic driving through Floquet protocols~\cite{claeys_floquetengineering_2019} and the creation of robust Hamiltonians~\cite{choi_robust_2020}. 

All these developments rely on the insight, that a system subject to a periodic drive can be described by an effective Hamiltonian, which may be related to the system Hamiltonian and simply feature renormalized couplings, but may also be completely different. However, there are strong arguments indicating that for generic (ergodic) interacting QMBS the periodic driving leads the system to heat up to infinite temperature resulting in a featureless state at late times~\cite{dalessio_longtime_2014,lazarides_equilibrium_2014}. The only exceptions known so far are many body localized systems, which are believed to resist heating to infinite temperature~\cite{ponte_periodically_2015, ponte_manybody_2015, lazarides_fate_2015, khemani_phase_2016} and an $O(N)$ model in the $N \to \infty$ limit \cite{chandran_interactionstabilized_2016}. While these arguments are not disputed in principle, over time a number of numerical studies have observed and reported absence of thermalization to infinite temperature in clean systems, which was either attributed to dynamical localization phenomena~\cite{dalessio_manybody_2013, ji_suppression_2018} or threshold behavior~\cite{haldar_onset_2018, heyl_quantum_2019, sieberer_digital_2019}, seemingly challenging the heating to infinite temperature paradigm. 

Another major challenge is the actual determination of the time it takes for a particular system of interest to heat up to infinite temperature. This question is particularly important for the Floquet engineering of e.g.~topological phases, where the prethermal regime governed by the effective Hamiltonian should be long enough to observe the transient stabilization of interesting phases, well before the heating dynamics takes over. Understanding the time scales in this setup has thus gained attention in the last years and there has been corresponding theoretical progress. Most significant perhaps are proofs that the heating time is exponentially large, typically $t_{h} \propto \exp(\omega / J) $ with some microscopic energy scale $J$,  in the high-frequency regime $\omega\gg J$ ($\hbar=1$ throughout the paper). The proofs use different analytical techniques such as an analysis of the errors in linear response theory \cite{abanin_exponentially_2015}, the Magnus expansion \cite{mori_rigorous_2016} or multiple rotating frame transformations \cite{abanin_effective_2017, abanin_rigorous_2017}. These approaches rely on bounds such as Lieb-Robinson bounds \cite{lieb_finite_1972, nachtergaele_liebrobinson_2010} and can typically not be used to obtain numerical estimates of the actual heating time in a specific system.
 It is therefore of significant importance to have an accurate, flexible, reliable and computationally efficient method at disposal to predict the heating times in driven quantum many body systems. 
 
So far only relatively few quantitative analyses of heating times (not necessarily focusing specifically on heating to infinite temperature) in Floquet systems have been performed. Most rely on explicit and computationally expensive real-time simulations of either sufficiently large finite-size systems~\cite{machado_exponentially_2019,  heyl_quantum_2019} or systems treated within truncated schemes such as Density Matrix Truncation \cite{ye_emergent_2020}, methods based on the Density Matrix Renormalization Group \cite{kollath_modulation_2006, kollath_spectroscopy_2006}, non-equilibrium Dynamical Mean Field Theory \cite{peronaci_resonant_2018, sandholzer_quantum_2019} or a Keldysh approach~\cite{weidinger_floquet_2017}. 
For effectively weakly driven systems, the Fermi Golden Rule (FGR) approach provides an accurate picture~\cite{mallayya_heating_2019}.
To our best knowledge, the latter one is the only studied method for generic systems, that is (significantly) computationally less expensive than the real-time simulations and applicable to generic systems but is restricted to effectively weak coupling. Part of the reason for the reduced cost is, that it can obtain accurate predictions from smaller systems than real-time simulations. The problem in working with small systems, as discussed in Refs.~\cite{dalessio_longtime_2014, seetharam_absence_2018, ji_suppression_2018} and also in later parts of this article, is that the Hilbert space is too small to support heating at large frequencies. This though does however not mean, that the information about heating time scales is not yet contained within small systems, as we will demonstrate here.
 
In this work, we analyze the appearance of avoided level crossings in the eigenvalues of the Floquet propagator and will be able to infer and quantitatively predict also very long heating times. The significance of avoided crossings in Floquet (and generally many-body) systems is well-known and documented in~\cite{holder_avoided_2005, eckardt_avoidedlevelcrossing_2008, bukov_heating_2016}, but to our best knowledge there have not yet been efforts to resolve individual (as we will see often very narrow) crossings systematically and to link them to the heating rate. 

The article is structured as follows: in Sec.~\ref{sec:floquet} we introduce some basic notions in Floquet theory necessary to follow the arguments in further sections. In Sec.~\ref{sec: heating qmbs} we review some of the prior work centered around heating in Floquet systems. In particular we explain how heating rates show in real-time simulations and how they are predicted using the FGR. Our method of predicting heating rates, based on avoided crossing spectroscopy, is described in Sec.~\ref{sec: heating numerics}, where we also make the connection to the FGR for weak drives. In Sec.~\ref{sec: driven chain} we discuss in detail the application of our method to a particular driven spin chain~\cite{heyl_quantum_2019,sieberer_digital_2019}. For this model we also identify certain commensurate parameter points, akin to discrete time crystals, where the effective Hamiltonian cannot be obtained by a perturbative expansion and show that our method detects these features and is still applicable and accurate. Finally in Sec.~\ref{sec: freq dep}, we use the driven spin chain with frequency dependent couplings, an important scenario in Floquet engineering, to illustrate the applicability of our method in this case as well. A different spin chain model~\cite{machado_exponentially_2019,ye_emergent_2020} with weakly broken spin inversion symmetry is discussed in Appendix~\ref{app: ye} as a further illustration of the power of our method.

\section{\label{sec:floquet}Elements of Floquet Theory} 

Floquet theory is concerned with the study of time-periodic quantum many body systems with Hamiltonian $H(t)$ with period $T = 2\pi / \omega$. A standard setup consists of an average Hamiltonian $H_{0}$ and a drive Hamiltonian $V$ as
 \begin{equation}
 \label{eq: def floq ham}
 H(t)=H_{0}+f(t)V,
 \end{equation}
 with a $T$-periodic function $f(t)=f(t+T)$ with zero mean.

 The propagator over a single period, formally given by the time-ordered exponential (with time-ordering operator $\mathcal{T}$)
 \begin{equation}
 \label{eq: def floq propagator}
 U(T)=\mathcal{T}\exp\left(-\iup \rmint_{0}^{T} H(t') \; \dup t'\right),
 \end{equation}
 has the eigenvalues $\lambda_i=\exp(-\iup \theta_{i})$. We call $\theta_i$ the eigenangles in the following.
 The Floquet Hamiltonian $\Heff$ is the generator of the propagator over one period
 \begin{equation}
 \label{eq: def Floquet ham}
 U(T)=\exp\left(-\iup T \Heff(T)\right)
 \end{equation}
  with eigenvalues $\theta_i/T$. It is not unique since the angles can be chosen modulo $2 \pi$. A common choice, that we also make, is to restrict the eigenangles to the first Floquet zone $ \theta_{i} \in (-\pi,\pi]$. 
 In this work, for simplicity, we focus on a square wave drive (also known as a switched or bang-bang protocol)
 
  \begin{equation}
 \label{eq: square wave}
 f(t)=\sign\left[\sin(\omega t)\right] = \frac{4}{\pi}\sum\limits_{m=0}^{\infty}\frac{\sin\left[(2m+1)\omega t\right]}{2m +1},
 \end{equation} which however is more naturally understood in a discrete sense using the product
 \begin{equation}
 U_\mathrm{sw}(\tau)=U_{-}U_{+}=\exp\left(-\iup \tau H_{-}\right)\exp\left(-\iup \tau H_{+}\right),
 \label{eq:protocol}
 \end{equation}
with the half period $\tau=T/2$ and the Hamiltonians $H_{\pm}=H_{0} \pm V$.
Such setups are often used in theoretical studies, since they can be simpler to analyze analytically and numerically. They also arise naturally in digital quantum simulation, for example via a Trotter decomposition of a time independent Hamiltonian, see e.g.~\cite{heyl_quantum_2019, sieberer_digital_2019}.

 \subsection{\label{ssec: eff ham} Floquet Hamiltonian}
 
The Floquet Hamiltonian, as defined in~\eqref{eq: def Floquet ham}, governs the stroboscopic evolution between periods. However, using another starting point within a period (i.e.~a different phase of the square wave), would lead to a different Floquet Hamiltonian. For this reason, notions such as Floquet or effective Hamiltonian are not always used to denote the generator of the evolution operator. Some literature rather reserves these names for a gauge invariant formulation, moving influences such as the initial phase to the so called kick operator~\cite{goldman_periodically_2014, bukov_universal_2015, eckardt_colloquium_2017}. We will not make use this formalism, but would like inform that the Floquet Hamiltonian as defined in~\eqref{eq: def Floquet ham} is not gauge invariant~\cite{goldman_periodically_2014}.  

The appearance of the Floquet Hamiltonian, which can have a non-trivial dependence on the average and drive Hamiltonians, is what makes Floquet systems an interesting research subject. Correspondingly, large efforts have been devoted to obtain approximations to the Floquet Hamiltonian, usually at large driving frequencies $\omega$~\cite{abanin_effective_2017, goldman_periodically_2014, bukov_universal_2015, eckardt_highfrequency_2015, holthaus_floquet_2015, mikami_brillouinwigner_2016, eckardt_colloquium_2017, rodriguez-vega_floquet_2018, vogl_flow_2019, vogl_analog_2019}. 
 In the digital setup~\eqref{eq:protocol} the Baker-Campbell-Hausdorff (BCH) series~\cite{reinsch_simple_2000} can be used at small half-period $\tau$ (high frequency $\omega$)
\begin{equation}
\label{eq: Heff bch}
T\Heff \approx TH_{0}+\iup\frac{T^{2}}{4}[V, H_{0}],
\end{equation} where we recognize that the average and the Floquet Hamiltonian correspond to each other to first order in $T$.

\begin{figure}
\centering
\includegraphics[width=0.999\linewidth]{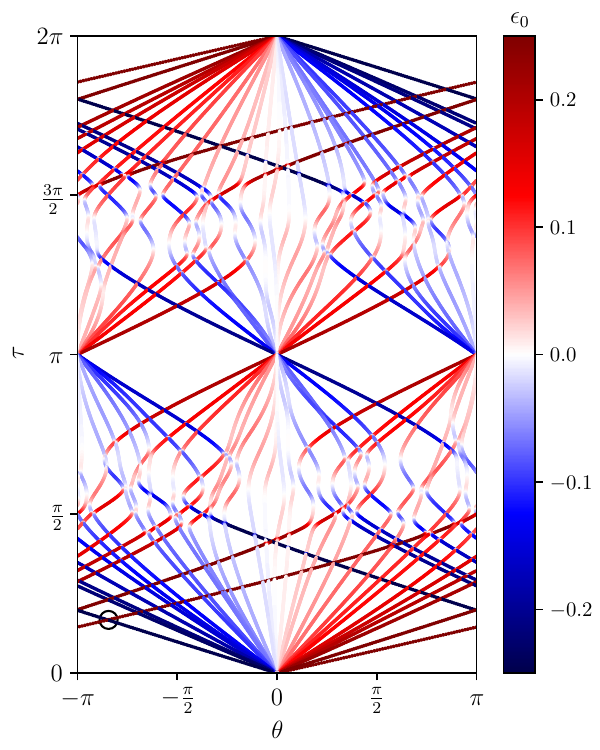} 
\caption{\label{fig:angles_colored} Floquet diagram of the spin chain with $ L=8 $ with eigenangles of $U_\mathrm{sw}(\tau)$ colored by their energy density with respect to $H_{0}$. The location of the first occurrence of an avoided crossing in this diagram is indicated by a black circle. Due to specifics of the protocol the eigenangles concentrate onto two points (one point) at $\tau=\pi$ (at $\tau=2\pi$).}
\end{figure}

\subsection{\label{ssec: floq diagram} Floquet Diagram}  
A visualization tool, used at times in the literature and providing a lot of insight for our method, is the Floquet diagram wherein the eigenangles $\theta_i$ of $U_\mathrm{sw}(\tau)$ for a finite size system are plotted as a function of the half-period $\tau$.  
For a model spin chain, to be specified in Sec.~\ref{ssec: hhz model}, such a diagram is shown in Fig.~\ref{fig:angles_colored}, where we only show the relevant symmetry sector. The lines denote the eigenangles $\theta_i(\tau)$. The color of the lines highlights the expectation value of the energy density with respect to the average Hamiltonian in the eigenstates of the Floquet propagator: $\epsilon_{0}=\langle \theta_i(\tau)|H_0/L|\theta_i(\tau)\rangle$.
Let us walk through some of the features of the Floquet diagram, which will be important in developing our method. At small $\tau$, starting from $\tau=0$, the lines
are almost straight as the Floquet propagator eigenstates and the eigenstates of $H_0$ basically coincide, and thus their slope in the diagram is proportional to the energy given by $H_0$. Incidentally, for the protocol family at hand~\eqref{eq:protocol}, this quantity is proportional to the derivative of the eigenangles with respect to $\tau$ at all values of $\tau$~\cite{claeys_breaking_2017}. Due to the $2\pi$ periodicity of the eigenangles, at a certain value $\tau{=}\tau_c$ the continuation of the lowest and highest energy state of $H_0$ seem to cross as indicated by a circle in Fig.~\ref{fig:angles_colored}. A more refined analysis shows that the two states undergo an {\em avoided} crossing with
a very small minimal angular gap of $\Delta\theta(\tau_c)\approx 10^{-7}\ \mathrm{rad}$. 

In a small $\tau{-}\theta$ region around this avoided crossing the many-body system can be sketched as an effective two-level system, whose dynamics can be understood as Rabi oscillations~\cite{gerry_introductory_2004} (see Appendix~\ref{app: rabi-osc} for an example). If we were to sit right on the crossing at $\tau_c$ and initialize the system in one of the two eigenstates of $H_0$, the off-diagonal matrix element responsible for the minimal gap would drive a (resonant) Rabi oscillation between the two $H_0$ eigenstates, therefore violating the energy conservation with respect to $H_0$ in an explicit manner. 
This exemplary first crossing thus provides an initial seed on a small system for the proliferation of many-body heating processes in larger systems. Our proposed method will build on this important intuition and consists of an automated analysis of {\em all} finite size level crossings in a certain window of $\tau$ and the $H_0$ energy transfer at each of them. 

Looking again at the Floquet diagram, we notice that for our specific choice of $H_0$ and $V$, the Floquet propagator $U_\mathrm{sw}(\tau)$ shows unusual behavior 
in the considered $\tau$ window at $\tau=\pi$ and $\tau=2\pi$, where the eigenangles join at $\theta=0,\pi$ and $\theta=0$ respectively. This behavior is closely related to discrete time crystals, where at least for $\tau=\pi$ one observes a recurrent dynamics with period $2\tau$ (i.e.~a period doubling) for generic initial states. While we relegate the discussion of the specific properties of the Floquet propagators at these values of $\tau$ to Sec.~\ref{sec: driven chain} and Appendix~\ref{app: hhz commensurate}, an important feature is that the number of avoided crossings as well as the magnitude of angular gaps, is strongly suppressed in the vicinity of those special points leading to a reduced heating rate, which competing methods such as the FGR treatment cannot easily access.

\section{\label{sec: heating qmbs} Heating in QMBS}

\subsection{\label{ssec: hhz model} Driven Spin Chain}

As an illustration for heating in QMBS and for our method in later sections, we focus on a model which was recently studied in the context of digital quantum simulation~\cite{heyl_quantum_2019, sieberer_digital_2019} and argued to exhibit 
a threshold behavior as a function of the half period $\tau$, i.e.~the absence of detectable heating below a threshold value of $\tau$. The model is defined by
\begin{equation}
H_{+}=X\equiv\sum_{i=1}^{L} s_{i}^{x},\quad H_{-}=Z+ZZ\equiv\sum_{i=1}^{L}s_{i}^{z}+s_{i}^{z}s_{i+1}^{z}\ ,
\label{eqn:HHZ}
\end{equation}
with spin one-half operators $s_{i}^{\alpha}$ 
and periodic boundary conditions~\footnote{The cited works use open boundary conditions, however, we verified a similar threshold behavior in our case.}, leading to
$$H_{0} =\frac{1}{2}(X+Z+ZZ), \quad V = \frac{1}{2}(X-Z-ZZ).$$
Any operator in the protocol has a spatial translation symmetry and a spatial reflection symmetry, which allows us to reduce the Hilbert space dimension by working in the zero momentum and even spatial parity sector throughout this article. This restriction is only possible for initial states lying fully within the given sector, for example translation invariant products states, which we use throughout the article. $H_0$ is an Ising model with transverse and longitudinal field, with parameter values not too far from other instances which have been reported to obey "eigenstates thermalization hypothesis" (ETH) properties~\cite{banuls_strong_2011, kim_testing_2014}. The average and the drive Hamiltonians can thus be characterized as generic (non-integrable) QMBS. 

\subsection{\label{ssec: heating phen} Phenomenology and Earlier Diagnostics of Heating}

Suppose that we evolve a pure state with the Floquet propagator $U_\mathrm{sw}(\tau)$ and monitor $\epsilon_{0}$ after each cycle. Since the system is a Floquet system with only discrete time translation invariance, the
average energy $\epsilon_0$ need not be conserved. The initial value is given by the expectation value of the average Hamiltonian in the initial (product) state. If the hypothesis of heating to infinite temperature holds true, then 
$\epsilon_0$ is supposed to approach zero at late times for our Hamiltonian (in the thermodynamic limit).

It has been predicted analytically \cite{abanin_exponentially_2015, abanin_theory_2016, abanin_effective_2017, abanin_rigorous_2017, mori_rigorous_2016} and observed in numerical simulations \cite{machado_exponentially_2019, mallayya_heating_2019, ye_emergent_2020}, that for large parts of the dynamics the energy density changes exponentially slowly $\epsilon_{0} \sim \exp(-\Gamma t)$ with the heating rate $\Gamma$, or equivalently the heating time $t_{h} \equiv 1 /\Gamma$, and that this time increases exponentially with the frequency of the drive.

\begin{figure}[htbp]
\includegraphics[width=8.5cm]{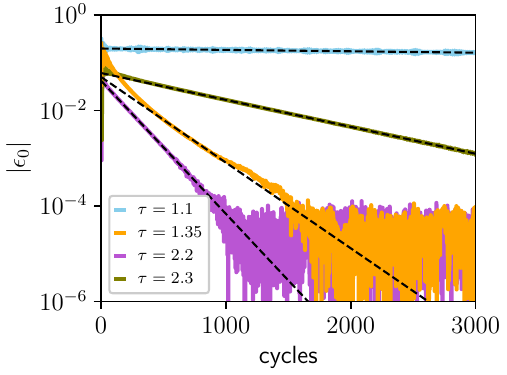}
\caption{\label{fig: energy_dynamics}
Real-time dynamics of the energy density of $\ket{x, +}$ for $L=28$ at different half-periods $\tau$. The energy density decays exponentially in time for large parts of the dynamics. For the smaller values of $\tau$ the heating time increases with frequency as expected. For larger $\tau$ this trend is reversed due to the specifics of the model.}
\end{figure}

 For certain small to intermediate system sizes, the heating rates can be obtained from real-time simulations using numerically exact methods. We perform the time evolution using Krylov subspace methods~\cite{higham_functions_2008} with partial reorthogonalization~\cite{simon_lanczos_1984} and appropriate error bounds~\cite{wang_error_2017} and show the results for a product state along the x-axis $\ket{x, +}$ in Fig.~\ref{fig: energy_dynamics} for four different values of the half-period $\tau$. 
In the figure we can clearly observe the almost perfect exponential decay of $|\epsilon_0|$, until the curves reach a finite size plateau with fluctuating $|\epsilon_0|$, here of the order of $10^{-4}$ for the system size of $L=28$ spins. 
In the thermodynamic limit the energy plateau would be at $0$, but due to the finite system size, it has a finite value decreasing with system size. More precisely, the steady state can be described as a random pure state, which can be inferred from computing the Shannon entropy $-\sum\limits_{i}P_{i}\ln P_{i}$ with probabilities in the computational basis. In a Hilbert space with large dimension $N$, this quantity is given by approximately $\ln(N) - 0.4228 $~\cite{wootters_random_1990}, which we also observe for the steady states with an error of $\approx 10^{-3}$. 

The heating times can be extracted by exponential fits in an appropriate window, which are indicated by black dashed lines. Even without extracting the rates, one can see clearly that the heating time decreases from $\tau=1.1$ to $\tau=1.35$ and $\tau=2.2$ as expected, but then the heating time {\em increases} again at $\tau=2.3$. In Sec.~\ref{sec: driven chain} we will explain this unusual behavior in more detail.

\begin{figure}[htbp]
\includegraphics[scale=1]{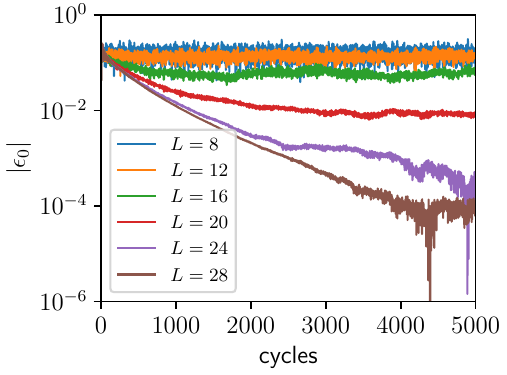}
\caption{\label{fig: heating finite-size} Dynamics of the energy density starting from $\ket{x, +}$ for various system sizes. At small sizes no heating is detectable, thus real-time simulations to extract the heating rate require large system sizes.}
\end{figure}

In complementary previous work the heating to infinite temperature was often diagnosed not from the actual real-time evolution of the energy, but instead through diagnostics which build on properties of the set of {\em all} eigenstates of the Floquet propagator $U_\mathrm{sw}(\tau)$. A prominent example is to determine the level spacing statistics of the eigenvalues of the propagator, or the (inverse) participation ratio of eigenfunctions~\cite{dalessio_longtime_2014, lazarides_equilibrium_2014, ponte_periodically_2015, heyl_quantum_2019}. These diagnostics build on the idea that for systems which heat up to infinite temperature the propagator is effectively an instance of a random unitary matrix (in the circular unitary (CUE) or circular orthogonal (COE) ensemble )~\cite{dalessio_longtime_2014,lazarides_equilibrium_2014}. In our work we demonstrate that these diagnostics are quite conservative, i.e.~they are typically unable to detect the heating to infinite temperature on system sizes which are too small in relation to the underlying heating time. If systems are too small, such that no heating is observed in real-time simulations, these measures also cannot be used to learn about heating for larger sizes~\cite{seetharam_absence_2018, ji_suppression_2018}. This is illustrated in Fig.~\ref{fig: heating finite-size}, where the dynamics of the energy density of $\ket{x, +}$ is shown for various system sizes. As seen in the figure, the smallest shown sizes do not seem to heat at all and one needs to go to $L \approx 24/28$ to really see a consistent heating rate. 

Our work based on avoided level crossings however directly focuses on the seeds of the heating processes, and is able to predict even very large heating times from rather small systems ($L \leq 12$), where the circular ensemble has not yet permeated most eigenstates of the propagator. Conversely, we will also see that our method is not well suited to extract heating times in regimes where the ensemble's properties are fully expressed, but since these are the "simple" cases, where the heating happens typically very fast, this is not an important limitation.

\subsection{\label{ssec: fgr} Fermi-Golden Rule}

As we will see shortly, there are certain parameter regimes wherein the heating rate changes over several orders of magnitude in a small $\tau / \omega$ window, rendering real-time simulations particularly expensive, since to resolve this region one needs to perform the evolution for multiple parameter values and potentially very long run times not known a priori.

In weakly driven systems, the FGR has been shown~\cite{mallayya_heating_2019} to give accurate predictions at a much lower cost. The FGR is rooted in time-dependent perturbation theory/linear response theory~\cite{gerry_introductory_2004, schwabl_quantenmechanik_2007, cohen-tannoudji_quantenmechanik_2019a} and proposes the following formula
\begin{align}
\label{eq:ear_fgr}
\dot{E}(\omega)=& \; \frac{\pi}{2}\sum\limits_{m}g^{2}_{m}\sum\limits_{i, j}\omega_{ji} |\braket{j|V|i}|^{2}P_{i} \nonumber \\
&\times \left[\delta\left(m \omega - \omega_{ji} \right)+\delta\left(m \omega + \omega_{ji} \right) \right],
\end{align}
for the extensive energy absorption rate (EAR) $\dot{E}(\omega)$ as a function of the driving frequency $\omega$. In the formula $g_m$ is the $m$-th Fourier component of the driving amplitude $f(t)$ and the set of $|i\rangle$ denotes the eigenstates of $H_0$ with energy $E_i$. Subsequently, $\omega_{ji}\equiv E_j-E_i$ is the energy difference between eigenstates and $P_i$ denotes the probability to find the system in eigenstate $|i\rangle$. The latter is needed since the heating rate is a priori state dependent, however during the evolution these probabilities of course change, which is not captured by the formula. 

In a numerical FGR computation one has to calculate all the matrix elements $\braket{j|V|i}$ of the drive Hamiltonian $V$ in the eigenstates of $H_0$, amounting to \emph{one} full diagonalization of $H_{0}$. Then using specific values of $\omega$, $g_m$ and a model for $P_i$ one can evaluate Eq.~\eqref{eq:ear_fgr}. 
Apart from providing a computation tool, the FGR also provides a way to understand the exponential increase of heating times at high frequency and sheds light on some statistical aspects of heating in many-body systems.

As a starting point one can rewrite the double sum as an integral over the density of states $\mathcal{D}(E)$ (details are discussed in the Supp. Mat. of Ref.~\cite{mallayya_heating_2019})
\begin{align}
\label{eq: fgr td limes}
\dot{E}(\omega)=& \; \frac{\pi}{2}\sum\limits_{m}g^{2}_{m}\rmint \dup E \;  \mathcal{D}(E)\mathcal{D}(E+m\omega) (m\omega) \nonumber \\
& \times |\braket{E+m\omega|V|E}|^{2}(P(E)-P(E+m\omega)).
\end{align}
Since the approach operates under the assumption that the system (i.e. average Hamiltonian) is generic, it is expected that the matrix elements of the drive are given by the ETH ansatz  
\begin{equation}
\label{eq: ETH fgr}
\braket{E+m\omega|V|E} \approx \frac{f_{V}(\bar{E}, \omega)}{\sqrt{\mathcal{D}(\bar{E})}}R,
\end{equation} 
where $\bar{E}$ is the average of $E, E + m\omega$, $R$ is a random variable with zero mean and unit variance and $f_{V}$ is a smooth function independent of system size~\cite{srednicki_chaos_1994, dalessio_quantum_2016}. 

For local operators $ O $, $f_{O}$ has been shown numerically to decay exponentially with $\omega$ for high frequencies in a variety of systems~\cite{beugeling_offdiagonal_2015, dalessio_quantum_2016, mondaini_eigenstate_2017}. This behavior can serve as an explanation for the exponential suppression of heating and threshold behavior from a statistical perspective. Furthermore, previous results from FGR (and our results from crossing computations) suggest that small systems can already provide good estimates for this function. In evaluating the formula for such systems, one effectively cancels the density of states factors and gets an estimate for the thermodynamic limit, where the behavior is dictated by $f_{V}$. 

\section{\label{sec: heating numerics} Heating Rates via Avoided Crossing Spectroscopy}

As we have discussed in the previous section, the success of the FGR is rooted in a sensible separation of microscopic processes embodied in $f_{V}$, from statistical factors like the density of states. We propose to construct the heating rate in a similar fashion, but using the true microscopic processes in the systems, encoded in avoided crossings, rather than the expression based on linear response theory. In later sections we will show that this method has clear advantages in several scenarios occurring in Floquet systems.

\subsection{\label{ssec: avoided crossing} Avoided Crossings}
Let us start by analyzing isolated avoided crossings in more detail. Examining the Floquet diagram from Fig.~\ref{fig:angles_colored}, we postulate that close to a crossing the Floquet Hamiltonian within the subspace of the two crossing states is

\begin{equation}
\label{eq: rabi model}
T\Heff^{(subs.)} = \delta(\tau) s^{z} + \Delta_{c} s^{x},
\end{equation}
where we note that the operators $s^\alpha$ do not act on the physical spins but are to be understood as acting on the states within the two-dimensional subspace of crossing energy levels. This model features an avoided crossing at $\delta = 0$, where the energy gap is $\Delta_{c}$~\footnote{The addition of a $s^{y}$ term does not change the main conclusions, provided that $\Delta_{c}$ is the total gap at $\delta=0$.}. The eigenstates are (anti-)symmetric superpositions of the up and down states (in the subspace). For large $\delta$ the eigenstates are essentially the up and down states, however which one of these is higher in energy depends on the sign of $\delta$. Going through the crossing the states switch, meaning that the up / down states have the energy of the opposite state before the crossing. 

In a QMBS the first crossings are the ones between the edge states, thus after the first crossing the effective Hamiltonian is the average Hamiltonian with the two outermost edge states switched. Hence, already at this point one needs an additional many-body operator in the Floquet Hamiltonian, leading to the breakdown of perturbative expansions. This also means that if the crossings are well separated in $\tau$, there is no energy absorption with respect to the average Hamiltonian (from their respective subspace) outside of the close vicinity of the crossing. At the crossings on the other hand, there are Rabi oscillations between the corresponding states resulting in "fast" dynamics, which might however still be very slow compared to the natural time scales of the total dynamics (see App.~\ref{app: rabi-osc}).

The dynamics for the two-level system can be obtained exactly and is a classic result~\cite{cohen-tannoudji_quantenmechanik_2019}. In case of an initial diagonal density matrix $\rho(0)=\mathop{\mathrm{diag}}\left(P_{0}(0), P_{1}(0)\right)$ the probability in the ground state is
\begin{equation}
\label{eq: dynamics single crossing}
P_{0}(t)=\frac{\Delta_{c}^{2}}{\delta^{2}+\Delta_{c}^{2}}\sin^{2}(\frac{\sqrt{\delta^{2}+\Delta_{c}^{2}}}{2T}t)(P_{0}(0)-P_{1}(0)).
\end{equation}
As we will argue later, these oscillations are the basis of FGR and our method, for which we need first to obtain some quantities for the individual crossings. 

Suppose we wanted to evaluate the formula in Eq.~\eqref{eq: dynamics single crossing} for a single crossing. For this we need $ T, \delta, \Delta_{c} $ and $P_{0}(0)-P_{1}(0)$, but since we are interested in the resonant oscillation we set $\delta=0$. This amounts to knowing the crossing time $\tau_{c}$, the angular gap width $\Delta_{c} $ and the pair of states $i, j$ which cross. To compute these quantities in practice, we first completely diagonalize the propagator $U_\mathrm{sw}(\tau)$ for many values of $\tau$. The $\tau$-resolution required depends on how narrow the gaps are in $\theta$ and how far apart they are in $\tau$. Currently, we work with
a fine uniform grid in $\tau$ and resolve the crossing locations and the minimal gaps for different grid resolutions. In future improvements this could also be done using an adaptive grid or automated root finding techniques. The determination of the energy transfer with respect to $H_0$ can be done in different ways. Here we chose for simplicity a scheme where we track pairs of crossing states back to their energy as $\tau\rightarrow 0$, this is done by working backwards through the crossing history of the involved states. Other possible ways to determine the energy transfer would be to measure the expectation values of $H_0$ in the pair of states before and after the crossing, or - for our particular protocol - to determine the slopes $\dot{\theta}(\tau)$ before and after the crossing. We leave these refinements for future work however.  Details of the algorithm(s) are discussed in Appendix~\ref{app: algorithm}.

In Fig.~\ref{fig: gap widths} we show the gaps for the driven chain with $L=8$ at two different resolutions in $\tau$. Here, one can see that the gaps change over several orders of magnitude in a small $\tau$ window. For
small values of the avoided gaps ($\lesssim 10^{-6}$) the resolution has a visible effect, however for the larger gaps there is not any noticeable difference.
 \begin{figure}[htbp]
\includegraphics[width=8.55cm]{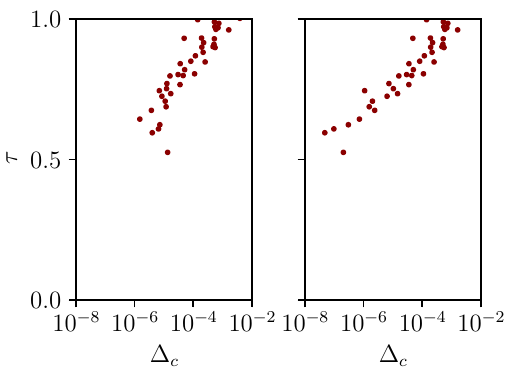}
\caption{\label{fig: gap widths} Gap widths of the driven chain with $L=8$ as obtained from the algorithm described at two different resolutions in $\tau$:  $\pi \cdot 10^{-6}$ (left) and $5 \cdot 10^{-8}$ (right).}
\end{figure}
  
\subsection{\label{ssec: Heating from Crossings} Heating from Crossings}

For deriving a formula for the EAR from a Rabi oscillation, we follow the linearization of the expression~\eqref{eq: dynamics single crossing} as in the derivation of the FGR~\cite{gerry_introductory_2004, schwabl_quantenmechanik_2007, cohen-tannoudji_quantenmechanik_2019a}. This amounts to using the identity
$$ \delta(\alpha)=\lim\limits_{t \to \infty} \frac{\sin^{2}(\alpha t)}{\pi \alpha^{2}t} ,$$
for the delta function, which leads to a linear rate rather than an oscillation.
Of course, this treatment can be valid only under certain assumptions, for example that the probability transfer is small, which are discussed in more details in the cited literature. A possible interpretation is that the dynamics can be viewed as an off-resonant (far-detuned) Rabi oscillation. 

Using the linearization procedure, we obtain the following formula for the EAR (details of the derivation along with a derivation of the FGR are laid out in Appendix~\ref{app: fgr})
\begin{equation}
\label{eq: ear crossings}
\dot{E}(\omega)=\frac{\pi}{2}\sum\limits_{m}\sum\limits_{c}\frac{\Delta^{2}_{c}}{T^{2}_{c}}\left(\Delta E_{c} \right)\left(\Delta P_{c} \right)\delta \left(m\omega - \omega_{c} \right),
\end{equation}
where the sum runs over all modes and the avoided crossings $c$ attributable to the corresponding mode, with $T_{c}$ the period, $\Delta_{c}$ the gap. $\left(\Delta E_{c} \right)$ is (the absolute value of) the energy difference of the states that cross with respect to the average Hamiltonian. In the FGR this would simply be given by the frequency $\omega$ (or multiples thereof). Generally one can use the difference in expectation values for any observable to obtain the absorption rate for that particular observable. Finally, $\left(\Delta P_{c} \right)$ is the difference in occupation of the crossing states. This of course depends on the instantaneous state of the system during the dynamics, however we will be using a high-temperature ansatz to obtain an estimate later.

Given the discussion above, the formula has an intuitive interpretation: at each avoided crossing transitions with the rate $(\pi/2)(\Delta_{c}/T_{c})^2 \delta \left( m\omega - \omega_{c} \right)$ occur and transfer an energy corresponding to the energy difference per unit time. The probabilities are a sort of "balancing" factor, such that transfer is enhanced for a large difference and vanishes in the fully mixed state (compare this to~\eqref{eq: dynamics single crossing}).

Presumably, the gap widths cannot be related directly to matrix elements in general. Therefore, an analysis of the convergence similar to the one in Sec.~\ref{ssec: fgr} will not be possible in general. However, in some cases, for instances at fast or strong drives, a relationship to matrix elements similar to the FGR can probably be recovered by transforming into an appropriate frame and applying the same formalism therein. Also as discussed in Sec.~\ref{ssec: driven chain commensurate} the Floquet formalism close to a discrete time crystal is very similar to the one at $\tau=0$, hence we expect the same convergence as in the FGR at this point.

Note as well, that even if a relationship with matrix elements is given, this does not imply that the sum over gaps in a finite system yields an accurate estimate for the thermodynamic limit. For example driven many-body localized systems have been reported to not heat in certain frequency windows~\cite{ponte_periodically_2015, ponte_manybody_2015, lazarides_fate_2015, abanin_theory_2016}. Clearly though, the naive evaluation of the formula for a finite, small to intermediate size, system would yield an observable rate. The lack of heating thus has to stem from a different scaling with system size of the number/width of the gaps compared to the FGR case and extracting this behavior would require a more in depth analysis than just the evaluation of the formula. Nevertheless, for ergodic systems or systems, possibly transformed to a suitable frame, the simple treatment can be justified and convergence with system size is expected.

This formula, along with the automated resolution of crossings, is the central result of our work. In the next paragraph, we will show that for weak drives it is equivalent to the FGR, but in subsequent sections it will also become clear that it has a much larger region of validity, since here the actual crossing in the concrete system are used instead of perturbative approximations.

\subsection{\label{ssec: compare fgr} Comparison with FGR}

Comparing formulas~\eqref{eq:ear_fgr} and~\eqref{eq: ear crossings} (note that the double sum in the FGR is actually a single sum due to the delta functions as well), we recognize that the FGR is a special case of the crossing based formula, wherein the crossing time as well as the energy transfer are given by the energy difference $\omega_{ij}$ between the states. This would be the case if the lines in the Floquet diagram were perfectly straight lines, which is reasonable for weak drives. Furthermore, the matrix elements and gap widths have to be related by 
$$ g^{2}_{m}|V_{ij}|^{2}=\frac{\Delta_{ij}^{2}}{T^{2}} \Leftrightarrow \frac{\Delta_{ij}}{T}=g_{m}|V_{ij}|,$$
which corresponds to the gap width one would obtain in the Rabi model in the two-state subspace as discussed in Appendix~\ref{app: fgr}. 

\begin{figure}[bp]
\includegraphics[width=8.55cm]{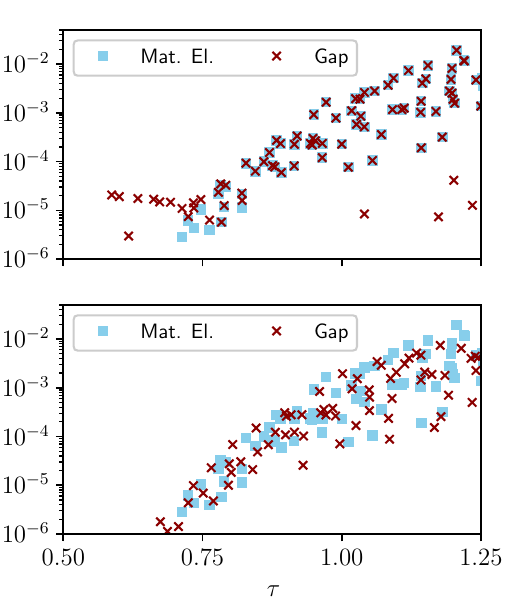}
\caption{\label{fig: matel} Matrix elements of the drive computed exactly and from gaps for the driven chain with $L=8$ and different drive strengths $g=0.01$ (upper) and $g=1$ (lower). For the weak drive, the data coincides very well apart from very small $\tau$, where the gaps are limited by resolution and few gaps likely from "multi-photon resonances". For the stronger drive, the magnitude fits reasonably well but the locations of the gaps are shifted.}
\end{figure}

We investigate this relation numerically by introducing a factor $ g $ for the drive strength, changing $V \to gV$ in the protocol. We then compare the exact matrix elements with the appropriate expressions from our computed gaps for the first mode ($g_{1}=4/\pi$ for the square drive). The results are shown in Fig.~\ref{fig: matel} for $L=8$ and drive strengths $g=0.01$ and $g=1$, where the latter corresponds to the original model.

Here we observe that for $g=0.01$ the correspondence is very good, apart from a region at very small $\tau$ where the gaps are limited by the resolution and some very small crossings which are likely due to "multi-photon resonances" i.e. levels with energy difference $\omega_{ij}$ meeting at frequency $\omega=m\omega_{ij}$ in the Floquet diagram (one can see this visually in Fig.~\ref{fig:angles_colored} - the levels with the largest slope meet a second time ($\omega_{ij}=2\omega$) within the $\tau-$window). At $g=1$ the expressions are still qualitatively similar, but especially the locations of the gaps are noticeably different. This means that strictly speaking the assumptions of the FGR are not valid anymore for the model i.e.~states do not cross at a $\tau$ given by the energy difference. However, for larger systems the density of crossings will increase and these small corrections will be washed out allowing the FGR to still make a good prediction. In this sense, in the rough region of validity of the FGR we expect our method to not improve predictions significantly. However, in the following sections we will show two scenarios in which our method, has clear advantages.

\section{\label{sec: driven chain} Driven Spin Chain}

\begin{figure*}[htbp]
\includegraphics[width=1\linewidth]{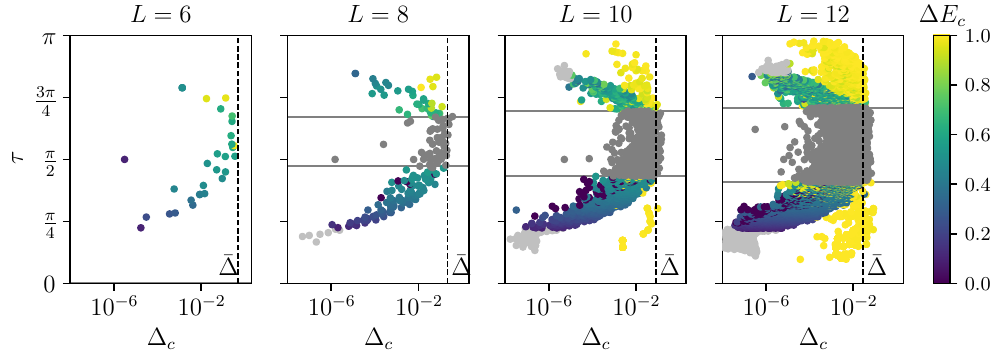}
\caption{\label{fig: energy transfer} Gap widths and energy transfer (color code) with respect to $H_{0}$ for different system sizes. The kink at small $\tau$ for larger sizes (light gray) is due to finite resolution in $\tau$ ($5 \cdot 10^{-8}$ for $\tau \leq 1$ and $\pi \cdot 10^{-6}$ otherwise). The results from expanding the states from $\tau=0$ and $\tau=\pi$ are shown in color, while in the central region (gray) the true energy transfers cannot be obtained by the expansion.}
\end{figure*}

Having described the method in general, let us now use the driven spin chain as a concrete example to demonstrate the power of our approach. The gap widths and energy transfers obtained as explained in Sec.~\ref{ssec: avoided crossing} are shown in Fig.~\ref{fig: energy transfer} for different even system sizes, ranging from $L=6$ to $L=12$.
On the $y$-axis we plot the half-period $\tau$ of the protocol. The points denote identified crossings, while the color scale of the points encodes the absolute value of the energy difference with respect to $H_0$  between the two Floquet eigenstates involved in the avoided crossing. Furthermore, we have indicated a dashed line at the mean angular spacing $\bar{\Delta}=2\pi/\dim \mathcal{H}$ dictated by the dimension $\dim \mathcal{H}$ of the relevant Hilbert space sector $\mathcal{H}$.

\subsection{\label{ssec: high freq part hhz} High-Frequency Region}
Let us first focus on smaller $\tau$ values $\tau \lesssim 2$: we observe that the magnitude of the gap widths increases over 4-5 orders of magnitude in this $\tau$ window. As the system size increases gaps at increasingly small values of $\tau$ appear as the spectrum now contains states with the corresponding frequency difference. For the larger system sizes some of these gaps are limited by the resolution resulting in blob like structures, which we color light gray. Furthermore, the magnitude at fixed $\tau$ remains roughly constant unless it is would be larger than the mean level spacing, which then acts as a cutoff for the magnitude of the gaps. We color the region where this is the case in darker gray. 

Finally, on a technical note, there seem to be some crossings with a magnitude and energy transfer that do not fit the overall picture. These occur for two reasons: first the algorithm as outlined in the previous section is very sensitive even to small wiggles between the distance of adjacent levels and therefore detects some "ghost" crossings even between levels that seem to evolve mostly straight. Since these "ghost" crossings are not accompanied by an actual swap of two states, these cause the ordering of switched levels during the algorithm to become inaccurate after a while. For a further discussion and illustration of the wrong order introduced by "ghost" crossings see Appendix~\ref{app: ye}. As a second reason, it turns out that many of the seemingly wrong crossings at $\tau \approx \pi$ are actually genuine crossings with a switching. The non-fitting magnitude and energy transfer here result from the fact, that levels originating at $\tau=\pi$ cross within the same subspace (take the two subspaces in Fig.~\ref{fig:angles_colored} as an example) and therefore have very similar energies. In principle one could correct for those by including some sort of curvature check in the algorithm to determine if a switching really took place or by simply discarding crossings where magnitude and energy transfer do not fit together. However, we found no noticeable effect of the non-fitting crossings, since they have a small energy transfer and there are relatively few of them (compared to "standard" crossings). Therefore, we move forward using the most straightforward scheme of the algorithm.

Let us finish the discussion by trying to understand the significance of the region, wherein the average angular level spacing due to the Hilbert space dimension is smaller than the gap width for small sizes (gray region in the figure) for our method, which is tied to some fundamental questions about heating in Floquet systems, particularly to the thermodynamic limit. To our best knowledge some of these questions have no definite answer yet, hence we give our best attempt at an interpretation of the results in the literature related to these questions in the context of our method. Clearly, in the thermodynamic limit the average level spacing vanishes and thus the information about microscopic processes as encoded in the gap widths is somehow hidden. The Floquet propagator then "has properties of matrices of the COE of random matrix theory" in the words of~\cite{dalessio_longtime_2014}. However, we still expect some structure depending on the frequency based on the results in~\cite{mori_rigorous_2016, abanin_effective_2017, abanin_rigorous_2017}, wherein the exponential timescale in the heating time at high frequency was established for many-body systems. This, along with the real-time simulations in current and other works (see references in Sec.~\ref{sec:introduction}) suggest that the EAR converges in the thermodynamic limit (as also discussed in Sec.~\ref{ssec: Heating from Crossings}). Hence, some of the structure visible at small sizes survives. In what form the information about the timescales enters the Floquet propagator for large systems is unclear to us. It might be that there are traces hidden in the spectrum, for instance there is a mechanism in the ETH leading to a "shrinking" of matrix elements with system size through the factor $1/\sqrt{\mathcal{D}(E)}$. However, it is doubtful, whether the ETH formalism can be applied at finite frequency for large sizes, because the eigenstates of the Floquet propagator are likely to be fully mixed in the basis of the average Hamiltonian at those sizes. Thus, the spectrum might also be (statistically) equal at all frequencies for large enough sizes. This latter scenario seems to be consistent with the results in~\cite{lazarides_equilibrium_2014, dalessio_longtime_2014}. Therefore, we can only operate under the assumption that the information we extract is indeed relevant for large sizes without proof. Staying within this assumption though, we see that smaller sizes have a larger frequency window, wherein the gap width is separated from the mean level spacing. However, having chosen a frequency, one should strive for the largest possible sizes, for which the gap width is still unaffected, because larger sizes lead to a much more accurate estimate for the density of states and a finer frequency resolution due to more available gaps, which are needed to compute a smooth curve for the energy absorption. 

\subsection{\label{ssec: driven chain commensurate} Commensurate Points}

Let us now focus on the upper half of the $\tau$ window from $\tau \approx 2$ up to $\tau = \pi$. Due to the discrete nature of the protocol and the commensurability of the coupling strengths in the Hamiltonian, the Floquet Hamiltonian is not simply chaotic for all small frequencies. Instead, at some frequencies the propagator resp. Floquet Hamiltonian take on simpler forms, which shows in the spectrum as the appearance of degenerate subspaces (in our case one or two - depending on system size). This effect can clearly not be captured by a perturbation theory based on the average Hamiltonian and thus is not captured by the FGR. A detailed analysis, carried out in Appendix~\ref{app: hhz commensurate}, reveals that at all integer multiples of $\pi$ the propagator takes on "simple forms" (but not always in the same way). At $\tau = \pi$ the system features a discrete time crystal \cite{yao_time_2018, else_discrete_2020} (albeit a fine-tuned one), since $U_\mathrm{sw}^{2}(\pi)=I$ and therefore there is no heating, but instead completely periodic dynamics with a doubled period.

In general such non-perturbative points at $\tau'$, where $T'\Heff(\tau')=H'$, can be integrated into the general formalism of Floquet expansions by expanding around $\tau'$. Writing $d\tau = \tau - \tau'$ and expressing the propagator as 
\begin{equation*}
  U(\tau)=\exp\left(-\iup H_{-}d\tau \right) \exp\left(-\iup H'\right)\exp\left(-\iup H_{+}d\tau \right),
\end{equation*}
the approximate Floquet Hamiltonian can again be obtained through the BCH series
\begin{equation}
T\Heff \approx H'+2d\tau H_{0}+2 \iup d\tau \frac{[V, H']}{2}.
\end{equation}
The radius of convergence is certainly more questionable for this expansion, and it might be more appropriate to transform into the rotating frame of $H'$ here. However, we do not make explicit use of the expansion and its only virtue is to show, how the average Hamiltonian appears away from the high-frequency limit. In fact the commutator term does not have matrix elements in the degenerate subspaces, therefore $H_{0}$ is responsible for the splitting to first order, irrespective of what $H'$ actually is. This can be observed in Fig.~\ref{fig:angles_colored} since $\Heff$ at $\theta$ equal to $\pi$ and $2\pi$ are different from one another. 

For the dynamics this means that close to the time crystal the dynamics is a combination of the fast (period 2 cycles) dynamics and the much slower heating. The point here is that our method detects this, as exemplified by the vanishing of the gaps in Fig.~\ref{fig: energy transfer} close to $\tau = \pi$, and therefore the gap widths and the crossing locations can be obtained without any changes to the algorithm. For our chosen heuristic to determine the energy transfer by tracing the crossing states back to their initial energy at $\tau=0$, we need to alter the reference point to $\tau=\pi$ in the regime close to $\tau=\pi$. This is however only a limitation of our simplistic heuristic, and a more robust determination of the energy transfer using previously mentioned ideas would not require a reference point to start with.

\subsection{\label{ssec: driven chain heating} Heating Rates}
We are now in a position to benchmark our avoided crossing spectroscopy method with large-scale real-time simulations as well as the FGR predictions for the driven spin chain. For the evaluation of Eq.~\eqref{eq: ear crossings} and Eq.~\eqref{eq:ear_fgr} we follow~\cite{mallayya_heating_2019} in using a high-temperature thermal state (usual Boltzmann distribution expanded to first order in $\beta$) as a model for $P_{i}$, as we expect the evolved state to be sufficiently mixed in the Hilbert space for large parts of the dynamics, and using a broadened delta function, for example a normalized Gaussian with width $dE$, mimicking the density of states in the thermodynamic limit (see also Appendix~\ref{app: algorithm}).

From the EAR the heating rate $\Gamma=1/t_h$ is obtained through
$$\Gamma = \frac{\dot{E}_{\beta}}{E_{\beta}-E_{\infty}}, $$
where the subscripts indicate the energy evaluated at high- and infinite-temperature. Using the high-temperature expansion, the resulting heating rate is independent of temperature and should
therefore give rise to a mono-exponential decay of the energy density towards zero. A more careful treatment would take into account the concrete occupations, which might be incorporated into a sort of Boltzmann equation using the ideas developed in this work, however we restrain from this here since our goal is to get a feeling for the time scales involved and especially to identify the region in $\tau$, wherein the heating time changes drastically as discussed in previous sections. 

\begin{figure}[bp]
\includegraphics[width=8.6cm]{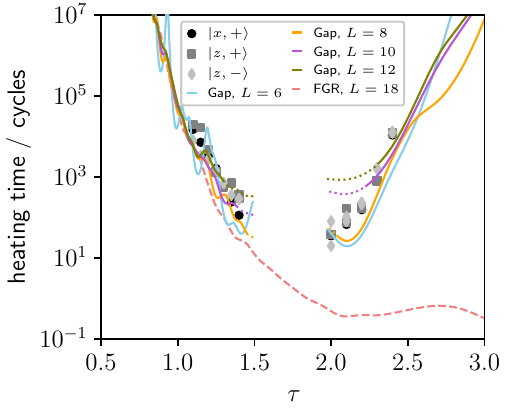}
\caption{\label{fig:heating_gap} Heating in the driven chain: Heating times of three states based on exponential fits to the energy (symbols). Predictions from FGR and based on the gap formula for different system sizes (lines). The dotted parts indicate the estimated range of validity for each system size. As discussed, for smaller $\tau$ the FGR and gaps agree well, while the FGR is unable to detect the commensurate point at $\tau=\pi$. Overall, both methods resolve a variation in the heating time of about five orders of magnitude, over a small change in half-period / frequency.}
\end{figure}

The heating rates obtained with the different methods (for $dE=0.1$) are shown in Fig.~\ref{fig:heating_gap}, which features heating times (measured in cycles) extracted from real-time simulations for three product states, the prediction based on FGR and the predictions based on avoided crossing spectroscopy for different system sizes. For the latter method we estimate a range of validity following the discussion in Sec.~\ref{ssec: high freq part hhz}.

The figure summarizes the earlier arguments, so let us also go through the main features again: for high frequencies the heating time increase rapidly and changes by several orders of magnitude, which is captured by both our method and the FGR. For lower frequencies, the FGR predicts a continuous decrease of heating times, while the observed times increase again due to the commensurate point at $\tau=\pi$. This is captured by the computed gap widths and using our method corresponding heating times can be extracted. The range of validity \emph{decreases} with the system size, hence smaller (and therefore computationally cheaper systems) can provide a more accurate estimate of the timescales involved. This comes however at the cost of featuring a lower density of crossings making it more difficult to obtain smooth curves for the heating times, if relying on broadening the delta function in the computation.

Overall, the proposed avoided crossing spectroscopy coincides with the FGR at high frequency, but also is accurate in resolving the temporal stability of the discrete time crystal as $\tau\rightarrow \pi$. It is impressive that the computation based on minimal assumptions such as the high-temperature ansatz and rather small systems sizes, ranging from $L=6$ to $L=12$, provide heating rates ranging over several orders of magnitude and capturing the regimes of rapid changes in the heating timescale very well. To increase accuracy one would need to improve on the model for occupations (the differences here are likely responsible for the different rates depending on the initial state) and to use different ways to introduce a density of states than to broaden the delta function for smaller sizes.

\section{\label{sec: freq dep} Systems with Frequency Dependent Couplings}

In the cases discussed above the average Hamiltonian played an important role, which could be understood within the expansion. If the coupling strength depends on frequency itself though, specifically if it diverges with frequency, the Floquet Hamiltonian is not necessarily given by the average in lowest order. This allows to simulate dynamics (within a given time scale) with a Hamiltonian that may otherwise be inaccessible and thus is an important tool in modern experiments (see references in the Introduction, Sec.~\ref{sec:introduction}). Note that oftentimes in the analysis of Floquet systems, no specific functional dependence of the frequency is specified a priori. Rather it turns out, that naturally a coupling strength $\propto \omega$ results in a sensible high-frequency limit. A well known example is the modification of tunneling in Bose-Einstein condensates~\cite{eckardt_superfluidinsulator_2005, eckardt_exploring_2009}. More recently, setups with strong couplings have been studied outside of the high-frequency limit and shown similar features~\cite{haldar_onset_2018, haldar_dynamical_2021}. As in the frequency independent case, different methods can be used to formulate high-frequency expansions for the effective Hamiltonian (see references in Sec.~\ref{ssec: eff ham}), which however often result in infinite series that cannot be summed analytically \cite{bukov_universal_2015}. Thus, we refer to the literature for the full details and content ourselves with a sketch of the argument using the BCH series here. 

We consider the switched setup from earlier, but make the drive strength proportional to the frequency $V \to(1 / \tau) V$. Therefore, the expressions $H_{\pm}\tau $ appearing in the propagator are given by $H_{0}\tau \pm V $. The BCH series consists of nested commutators, including commutators of the form $[\hdots[H_{0}\tau, V], V], \hdots V]$ (or other orderings), which, different from the independent case, are all $O(\tau)$ and thus contribute to the Floquet Hamiltonian. Hence, $H_{0}$ is only one of the (typically infinitely many) terms at the lowest order. Also, the terms can introduce interactions of all ranges and lead to very complex Hamiltonians, even from basic ingredients. In the remainder of this section, we will show that our method can provide useful results even in this setup, when neither the effective Hamiltonian nor the effective drive is available \footnote{We call here the new operator that is responsible for avoided crossings effective drive for lack of other terminology.}. 

\subsection{\label{ssec: driven chain freq model} Driven Chain with Frequency Dependent Couplings}

\begin{figure}[bp]
\includegraphics[width=8.6cm]{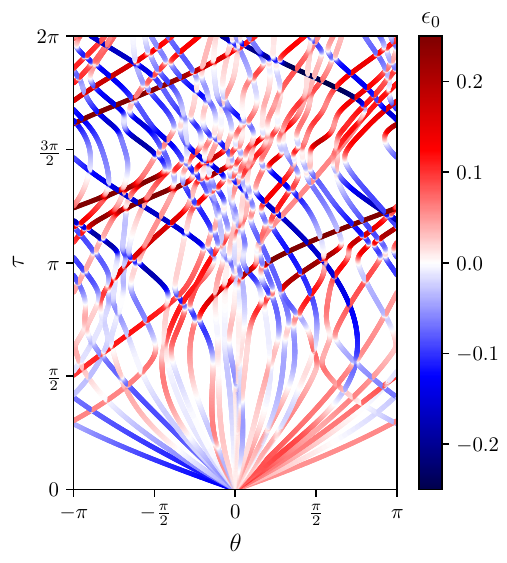}
\caption{\label{fig: colored freq} Eigenangles of the frequency dependent driven chain with $L=8$ colored by the (average) energy density. Compared to the frequency independent case we observe that the angles are not ordered strictly by energy at small $\tau$, show larger curvature and do not join at any (finite) values of $\tau$.}
\end{figure}

In order to illustrate the effect of frequency dependence, we stay with the driven spin chain from previous sections and modify it slightly $V \rightarrow (4 / \tau)V$, where the factor of four is chosen such that the additional terms have visible effects, but are not strong enough to change the overall scales significantly, such that we can operate in the same $\tau$ windows as before. The main conclusions concerning the applicability of our method are however not dependent on this choice, as will be apparent from the discussion. In Fig.~\ref{fig: colored freq} we show the eigenangles colored by $\epsilon_{0}$, where we can see that the slopes are not governed by $\epsilon_{0}$ and also display a stronger curvature overall. Also, the commensurate Floquet points at $\tau = \pi, 2\pi$ vanish as expected.

\begin{figure}[tp]
\includegraphics[width=8.6cm]{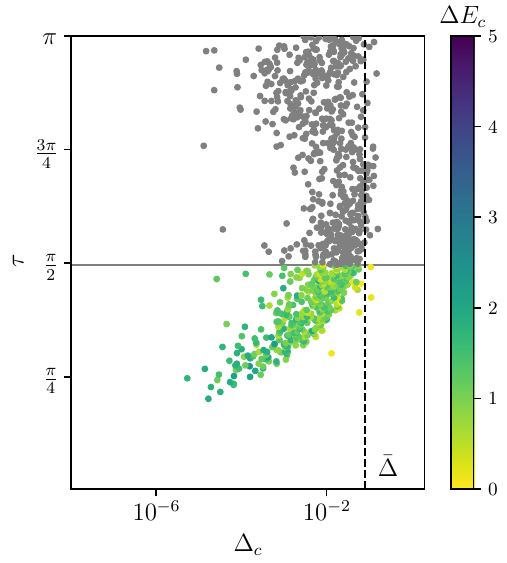}
\caption{\label{fig: transfer freq} Energy transfer of the frequency dependent driven chain with $L=10$. Compared to the frequency independent case the transfers do not correspond to the frequency and the shape of the decay region appears somewhat changed.}
\end{figure}

In Fig.~\ref{fig: transfer freq} we show the energy transfers with respect to $H_{0}$ obtained again by tracking the switchings between states. The overall behavior of the gap widths at high frequency looks similar to the frequency independent case, therefore we verify additionally that the new Floquet Hamiltonian is in fact significantly different from the average in Appendix~\ref{app: hhz freq}. This is also visible from the mismatch between the energy transfers and the frequency of the drive, in contrast to the frequency independent case.  

\subsection{\label{ssec: driven chain freq heating} Heating Rates}
\begin{figure}[htbp]
\includegraphics[width=8.6cm]{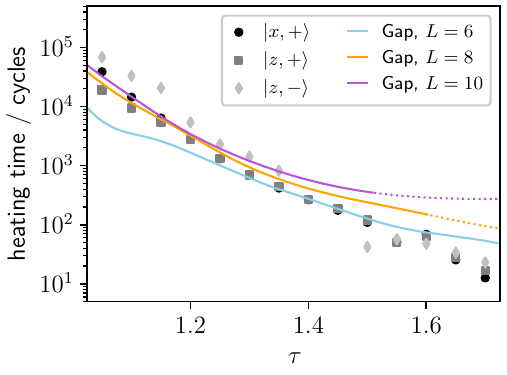}
\caption{\label{fig: heating freq} Heating times in the frequency dependent driven chain obtained from real-time simulations using three states with $L=26$ (symbols) and the gap data for different system sizes (lines). The dotted parts indicate the estimated range of validity for each system size. The agreement overall is reasonable, with different sizes showing good agreement in different parts as discussed in the text.}
\end{figure}

In Fig.~\ref{fig: heating freq} we finally show the estimated heating rates, evaluated with $dE=0.3$, as well as extracted rates from real-time simulations. Again, the dotted parts indicate the estimated range of validity for a given size.

As in the frequency independent case the agreement is reasonable overall, while being better at lower frequencies for the smaller sizes and better at higher frequencies for the larger sizes. Furthermore, in the Floquet diagram shown in Fig.~\ref{fig: colored freq} we observe that the energy (color code) seems to change during the evolution, as at the bottom there are no saturated levels while at intermediate $\tau$ there is some saturation. Hence, the extraction of energy transfers based on the original values also potentially leads to a lower accuracy.

\section{\label{conclusion} Conclusion and Outlook}
In this article we have shown how to analyze isolated avoided crossings in Floquet systems and how one can construct a versatile and accurate estimate for heating times based on those crossings. We have discussed that this method is closely related to the FGR, but with the demonstrated potential to go beyond it, since the crossings include non-perturbative effects. For this we have given two concrete examples using a driven spin chain: a discrete time crystal for commensurate points in a digital Floquet setting and a Floquet Hamiltonian beyond the average Hamiltonian due to frequency dependent couplings. In Appendix~\ref{app: ye} we have also shown that the setup can be used to detect non-generic behavior (here weak symmetry breaking) in a seemingly generic system. Furthermore, throughout the paper we have discussed how the method combines microscopic and statistical aspects and how this understanding can be used to understand why the method performs well in small systems and to obtain estimates for the region of validity at at given system size.

The approach introduced in this paper has the potential to address and potentially solve long-standing issues, such as the detailed heating dynamics in driven Bose and Fermi-Hubbard systems, where multiply occupied sites seem to have slow dynamics, and we also believe avoided level-crossing spectroscopy in an adapted form is able to shed light on the intricate relaxation and thermalization dynamics of non-integrable quantum many body systems, such as the quenched Bose-Hubbard model~\cite{kollath_quench_2007, biroli_effect_2010}.

For most of the numerical computations and the creation of the figures we use Python~\cite{langtangen_primer_2009} with the packages~\cite{hunter_matplotlib_2007, vanderwalt_numpy_2011, oliphant_guide_2015, lam_numba_2015, virtanen_scipy_2020}.
The data for all figures, as well as corresponding plot scripts, are freely accessible online~\cite{rakcheev_dataset_2020}.

\begin{acknowledgments}
The authors acknowledge support by the Austrian Science Fund FWF within the DK-ALM (W1259-N27). The computational results presented here have been achieved (in part) using the LEO HPC infrastructure of the University of Innsbruck. 
\end{acknowledgments}
\appendix
\section{\label{app: rabi-osc} Rabi-Oscillation Example}

In the main paper we discussed that at individual crossings the eigenstates of the Floquet Hamiltonian in the subspace are the (anti-)symmetric superpositions of the original states, which therefore perform a Rabi oscillation, with the probabilities oscillating as
\begin{equation}
P(t) \sim \sin^{2}\left( \frac{\Delta_{c}}{2 T}t \right),
\end{equation}
where $\Delta_{c}$ is the gap width at the crossing. We verify this for two distinct crossings, by identifying the gap width, gap position and the crossing states using the methods discussed in the previous section. As we will see, the resonance region is very narrow, hence for these specific crossings we manually improve on the exact values. 

\begin{figure*}[htbp]
\subfloat[$\tau=0.5249327354$]{\label{fig: rabi1 1} \includegraphics[scale=0.67
]{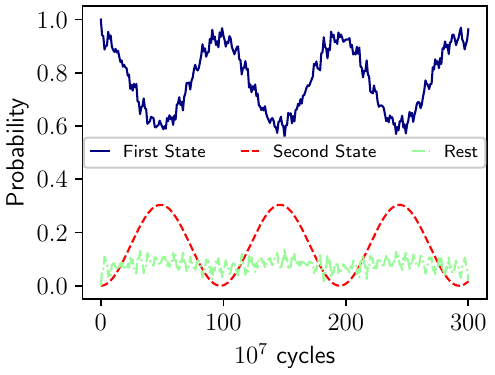}}
\hfill
\subfloat[$\tau=0.5249327359$]{\label{fig: rabi1 2} \includegraphics[scale=0.67
]{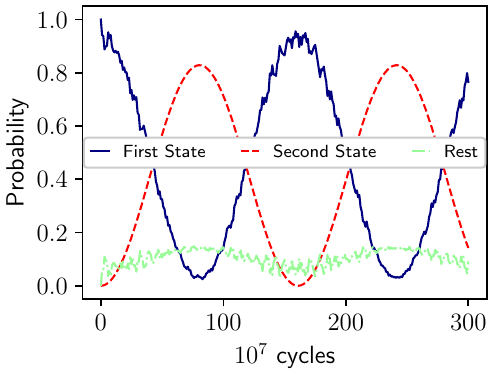}}
\hfill
\subfloat[$\tau=0.5249327364$]{\label{fig: rabi1 3} \includegraphics[scale=0.67
]{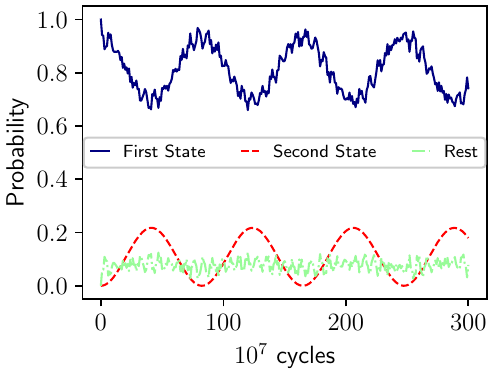}}
\caption{\label{fig: rabi osc 1} Rabi oscillation for the spin chain with $L=8$ starting from the lowest energy eigenstate of $H_{0}$. The probabilities are in the basis of $H_{0}$. The oscillation period matches the value obtained from the gap width $\Delta_{c} \approx 3.7 * 10^{-9}$. The contribution from other states results from perturbative corrections to the eigenstates.} 
\end{figure*}

 The results for the first (highest frequency) crossing can be seen in Fig.~\ref{fig: rabi osc 1}. In the figure we show the dynamics of the lowest energy eigenstate of $H_{0}$ for three different values of $\tau$ (before, at and after the resonance). One can see clearly that the dynamics is restricted to the subspace of the lowest and highest energy state to a large degree. It is not fully in the subspace, since outside of the vicinity of the crossing, the eigenstates of the Floquet Hamiltonian are eigenstates of $H_{0}$ with perturbative corrections. Given that to obtain the heating rate one needs to sum over all states in a given window, this supports the argument that energy absorption is governed by off-resonant oscillations.

\section{\label{app: fgr} Fermi Golden Rule and Energy Absorption}
 The Fermi Golden Rule can be derived from time-dependent perturbation theory. In this Appendix we present the main steps in the derivation, mostly following~\cite{schwabl_quantenmechanik_2007}, and then show, how these ideas can be used to derive Eq.~\eqref{eq: ear crossings}. Finally, the relationship between the gap widths and the matrix elements discussed in Sec.~\ref{ssec: compare fgr} 
 will be justified.
 
 In the derivation of the FGR, we are first concerned in the transitions between eigenstates $\ket{n}$ of $H_{0}$ under an evolution generated by $H(t)=H_{0}+\sum\limits_{m>0}g_{m}\sin(m \omega t)V$, with $g_{m} \in \mathbb{R}$. In the interaction picture, the propagator can be approximated by the first term in the Dyson series
 \begin{equation}
 U(t) \approx I - \iup\rmint\limits_{0}^{t}\eup^{\iup H_{0}t'}V(t')\eup^{-\iup H_{0}t'} \; \dup t',
 \end{equation}
 and our goal is to compute the transition probability $P_{nk}(t)=|\braket{k|U(t)|n}|^{2}$. Some steps can be performed exactly
 \begin{align}
 \label{eq: P_nk fgr}
 &P_{nk} = \bigg|\rmint\limits_{0}^{t}\eup^{\iup \omega_{kn}t'}V_{kn}\sum\limits_{m>0}g_{m}\sin(m \omega t) \; \dup t'\bigg|^2 \nonumber\\
 &=|V_{kn}|^{2}\bigg|\sum\limits_{m>0}\frac{g_{m}}{2\iup}\rmint\limits_{0}^{t}\left(\eup^{\iup(\omega_{kn}+m\omega)t'}-\eup^{\iup(\omega_{kn}-m\omega)t'} \right) \; \dup t'\bigg|^{2} \nonumber\\
 &=\frac{|V_{kn}|^{2}}{4}\bigg|\sum\limits_{m>0}g_{m}\left(\frac{\eup^{\iup(\omega_{kn}+m\omega)t}-1}{\omega_{kn}+m\omega}-\frac{\eup^{i(\omega_{kn}-m\omega)t}-1}{\omega_{kn}-m\omega}\right)\bigg|^{2}, 
 \end{align}
 where $\omega_{kn}=E_{k}-E_{n}$ and $V_{kn}=\braket{k|V|n}$. If we were to expand the absolute value, we would get a double sum over modes with "mixed" and "diagonal" terms. Within the diagonal terms, there are also "mixed" terms stemming from different denominators. It can be argued~\cite{schwabl_quantenmechanik_2007} that the contributions from the "mixed" terms can be neglected, and the remaining expression is
 \begin{align}
 P_{nk}(t)\approx& \; \frac{|V_{kn}|^{2}}{4}\sum\limits_{m>0}g^{2}_{m}\bigg(\frac{\sin^2((\omega_{kn}+m\omega)t/2)}{((\omega_{kn}+m\omega)/2)^2} \nonumber\\
 &+\frac{\sin^2((\omega_{kn}-m\omega)t/2)}{((\omega_{kn}-m\omega)/2)^2} \bigg),
 \end{align}
 where the identity $|\eup^{\iup xt}-1|^{2}=4\sin^{2}(xt/2)$ was used. The expression can now be "\textit{linearized}" using the representation
 \begin{equation}
 \delta(\alpha)=\lim\limits_{t \to \infty} \frac{\sin^{2}(\alpha t)}{\pi \alpha^{2}t} 
 \end{equation}
 for the Delta function. Inserting this yields 
 \begin{equation}
 P_{nk} \approx \frac{\pi}{2}|V_{kn}|^{2}\sum\limits_{m>0}g^{2}_{m}\left(\delta(\omega_{kn}+m\omega)\\+\delta(\omega_{kn}-m\omega)\right)t,
 \end{equation}
 and we can define the transition rate $\Gamma_{nk}=\dot{P}_{nk}$
 \begin{equation}
 \Gamma_{nk}=\frac{\pi}{2}|V_{kn}|^{2}\sum\limits_{m>0}g^{2}_{m}\left(\delta(\omega_{kn}+m\omega)+\delta(\omega_{kn}-m\omega)\right).
 \end{equation}
 We now consider a state with occupations $P_{n}$ and determine the change in energy due to transitions with rates $\Gamma_{nk}$. Each transition has an energy transfer rate $(E_{k}-E_{n})\Gamma_{nk}=\omega_{kn}\Gamma_{nk}$ and the total energy absorption rate is given by
 \begin{align}
 \dot{E}=& \; \sum\limits_{n}P_{n}\sum\limits_{k}\omega_{kn}\Gamma_{nk}=\frac{\pi}{2}\sum\limits_{m}g^{2}_{m}\sum\limits_{n, k}P_{n}\omega_{kn}|V_{kn}|^{2}\nonumber\\
 &\times \left(\delta(\omega_{kn}+m\omega)+\delta(\omega_{kn}-m\omega)\right).
 \end{align}
 This is precisely the FGR as stated in Eq.~\eqref{app: fgr}. We now apply some further manipulations to get the formula to a form closer to Eq.~\eqref{eq: ear crossings}. For this we first consider the exchange of indices $n, k$ in the sum: $P_{n} \to P_{k}, \;  \Gamma_{nk} \to  \Gamma_{kn}= \Gamma_{nk},\; \omega_{kn} \to  \omega_{nk}= -\omega_{kn}$. We therefore can rewrite the sum as 
 \begin{align}
 \dot{E}=& \; \frac{\pi}{2}\sum\limits_{m}g^{2}_{m}\sum\limits_{n > k}\omega_{kn}(P_{n}-P_{k})|V_{kn}|^{2} \nonumber\\
 &\times \left(\delta(\omega_{kn}+m\omega)+\delta(\omega_{kn}-m\omega)\right).
 \end{align} 
 Finally, we recognize, that due to the delta functions only terms with exactly matching energies contribute, hence the double sum is a single sum in disguise, and we can write this as
   \begin{align}
 \dot{E}=& \; \frac{\pi}{2}\sum\limits_{m}g^{2}_{m}\sum\limits_{\omega_{kn}=\pm m \omega}\omega_{kn}(P_{n}-P_{k})|V_{kn}|^{2} \nonumber \\
 &\times \left(\delta(\omega_{kn}+m\omega)+\delta(\omega_{kn}-m\omega)\right).
 \end{align} 
 Comparing this to Eq.~\eqref{eq: ear crossings}, we recognize that both coincide, given that $\Delta E_{c}=m|\omega_{kn}|, \; \Delta P_{c}=P_{n}-P_{k}$ and $\frac{\Delta^{2}}{T^{2}}=g^{2}_{m}|V_{kn}|^{2}$. Note that here we order the states such that $\omega_{kn}>0$ and thus only one Delta function is included. Furthermore, the sum over all avoided crossings implicitly includes the sum over modes, since in the weak drive (FGR) regime the levels with $\omega_{kn}=m\omega$ meet at $T=2\pi/\omega$ as discussed in Sec.~\ref{ssec: floq diagram}. Therefore, the avoided crossings include contributions from all modes. 
 
 Having derived the FGR from perturbation theory, we now consider deriving Eq.~\eqref{eq: ear crossings} from the dynamics of isolated avoided crossings. We begin though, by briefly recalling the main results from the Rabi model~\cite{gerry_introductory_2004, cohen-tannoudji_quantenmechanik_2019a}, using which we can make a connection between the effective Hamiltonian in the subspace (Eq.~\eqref{eq: rabi model}) and matrix elements in a weak drive limit. For this we look at the dynamics of a two-level system under a single mode drive described by the Hamiltonian
 \begin{equation}
 H(t)=H_{0}+gV\cos(\omega t).
 \end{equation}
 This Hamiltonian can be solved exactly within the "\textit{rotating wave approximation}". The solution for the transition probability is~\cite{cohen-tannoudji_quantenmechanik_2019a}
 \begin{align}
 P_{nk}(t)=& \; \frac{g^{2}|V_{nk}|^{2}}{g^{2}|V_{nk}|^{2}+(\omega - \omega_{nk})^{2}} \nonumber \\
 &\times\sin^{2}\left(\frac{\sqrt{g^{2}|V_{nk}|^{2}+(\omega - \omega_{nk})^{2}}t}{2}\right).
 \label{eq: rabi osc full}
 \end{align}
 From this expression, Eq.~\eqref{eq: P_nk fgr} can be obtained by taking the high detuning limit $(\omega-\omega_{nk}) \gg g_{m}|V_{ij}|$
 \begin{equation}
   \label{eq: rabi osc high detuning}
  P_{nk}(t)=\frac{g_{m}^{2}|V_{nk}|^{2}}{(\omega - \omega_{nk})^{2}} \sin^{2}\left(\frac{(\omega - \omega_{nk})t}{2}\right).
 \end{equation}
 We now compare the full solution to the dynamics in the static model $H=\frac{\delta}{T}s^{z}+\frac{\Delta}{T}s^{x}$,
 where $s^{x}$ could be replaced by a combination of $s^{x}$ and $s^{y}$ with the same spectrum. The solution reads~\cite{cohen-tannoudji_quantenmechanik_2019}
 \begin{equation}
 P_{nk}(t)=\frac{\Delta^{2}}{\Delta^{2}+\delta^{2}}\sin^{2}\left(\frac{\sqrt{\Delta^{2}+\delta^{2}}t}{2T}\right).
 \end{equation}
 Comparison to Eq.~\eqref{eq: rabi osc full} shows that in the weak drive limit $\frac{\Delta}{T}=g_{m}|V_{nk}|$ and $\frac{\delta}{T}=m\omega - \omega_{nk}, $ provided that the resonance of the $m-$th mode is targeted. Of course the entire derivation assuming one mode and a fully decoupled subspace is not strictly valid, however in the regime with very small gaps the levels are well isolated and the energy differences are reasonably large for the resonances from different modes to be well separated. 
 
 Finally, let us consider the off-resonant (high detuning) limit off an avoided crossing characterized by $\omega_{c}$ and $\Delta$, where as above we identify $\delta/T=m\omega - \omega_{c}$. Note that here we dot necessarily associate $\omega_{c}$ with the energy difference between the states, allowing for more general scenarios, such as the discrete time crystal point discussed in Sec.~\ref{ssec: driven chain commensurate}. The transition probability then reduces to
 \begin{equation}
 P_{nk}(t)=\frac{\Delta^{2}}{T^{2}(m\omega - \omega_{c})^{2}}\sin^{2}\left((m\omega - \omega_{c})\frac{t}{2}\right).
 \end{equation}
 Following exactly the same procedure as described before, we arrive at Eq.~\eqref{eq: ear crossings}.
\section{\label{app: algorithm} Details of the Algorithms}
Here we summarize how the gaps are computed, how the energy differences are then extracted and how Eq.~\eqref{eq: ear crossings} is evaluated. In the text we sketch the corresponding algorithms, with concrete code snippets in \textit{Python} being available in~\cite{rakcheev_dataset_2020}. 

\begin{figure*}[htbp]
\includegraphics[width=\textwidth]{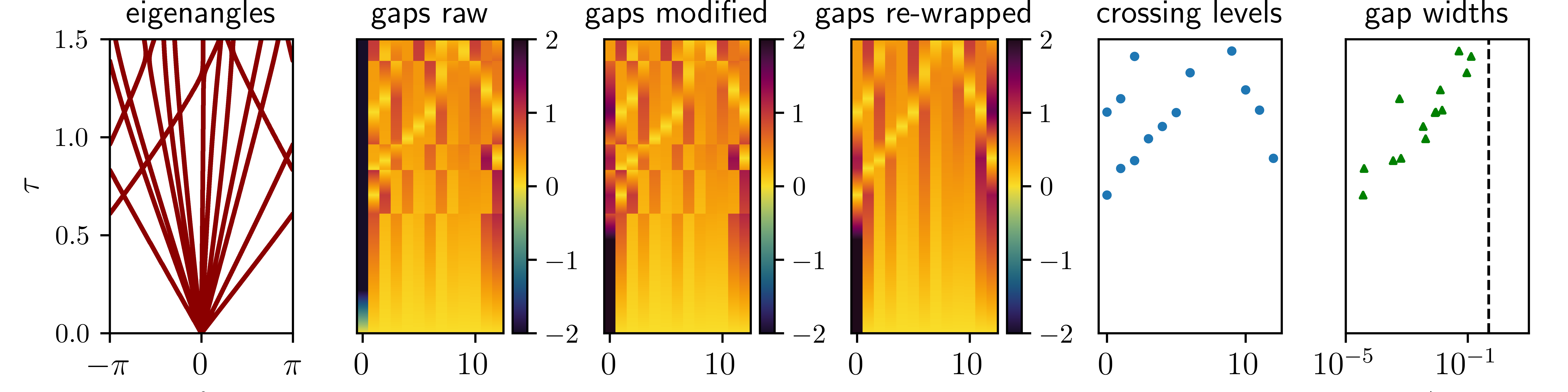}
\caption{\label{fig: gap finder} Illustration of the gap finding algorithm - see the text in Appendix~\ref{app: algorithm} for an explanation.}
\end{figure*}

We assume that we have an array of eigenangles $\theta_{i}(\tau_{n})$ (in the first Floquet zone $-\pi \leq \theta < \pi$) for a grid of half-periods $\tau_{n}$. In the following we sketch the main steps, which are also illustrated in Fig.~\ref{fig: gap finder}. 
\begin{enumerate}
\item We start with the eigenangles on a grid (leftmost subfigure).
\item First the angles are sorted in ascending order at each half-period, then we compute the differences between consecutive levels $\Delta_{i}=\theta_{i}$ at each $\tau$ including the first and last level (second subfigure from left).
\item The difference between the first and last level ($\Delta_{0}$ in the figure) has a redundant factor of $-2 \pi$, which we compensate for by adding $2 \pi$ to it at each $\tau$ (third subfigure).
\item There are discontinuities in the values of $\Delta_{i}$ due to levels wrapping around the first Floquet zone and distorting the ordering. We follow the levels at every $\tau$ step (starting from $0$) and check if the mean difference between consecutive slices is too large. For this we use the change in the previous step and compare this to the change in the current step. If the change is too large (we define work with a threshold of ten times larger) we "re-wrap" the next slice until it fits (fourth subfigure). This requires a good enough resolution, which is however needed anyways to resolve small gaps.
\item We can now use a local minima search to identify the locations of the minimal gaps. In the two last subfigures we show the resulting crossing locations i.e. which levels cross and the gap widths together with the crossing times.
\end{enumerate} 

Having this information we can identify the energy differences at each crossing by backtracking (here to the origin). For this we simply need the spectrum of $H_{0}$ sorted in ascending order. We now go through the crossings in order of $\tau$, noting the energy difference of the crossing levels as well as the switching in the spectrum due to the crossing. For instance the first crossing is between the first and last state, so we note the energy difference of these states, switch the first and last entry in the spectrum and then proceed to the next crossing.

Finally, to compute the EAR we need to evaluate Eq.~\eqref{eq: ear crossings}. For this we replace the delta function by a (normalized) Gaussian with width $dE$ We use $dE \approx 0.1$ in the examples (check the corresponding sections for concrete values), which we arrive at by starting from low values and increasing until a more or less smooth curve emerges. Also, as discussed earlier, we use a high-temperature ansatz for $\Delta P_{c}$, wherein $\Delta P_{c} = |\eup^{-\beta E_{i}} - \eup^{-\beta E_{j}}|/Z(\beta) \approx \beta|\omega_{ij}|/ \dim \mathcal{H}$ with $i, j$ referring to the crossing levels. Substituting this into the definition of the heating rate 
$$\Gamma = \frac{\dot{E}_{\beta}}{E_{\beta}-E_{\infty}}, $$
with $E_{\beta}$ being the thermal expectation value and $E_{\infty}$ the expectation value at infinite temperature, one gets an expression independent of $\beta$. In practice, we observe that at $\beta \approx 0.001$ the result is well in the regime independent of $\beta$ and thus use this value as default. A discussion of the ansatz can also be found in the Supp. Mat. of Ref.~\cite{mallayya_heating_2019}. For larger sizes this procedure is more stable and leads to similar result over a range of widths, while smaller sizes are more sensitive due to the low density of crossings. 
\section{\label{app: hhz commensurate} Propagator at Commensurate Points}

We want to understand why the Floquet propagator takes on simple forms at integer multiples of $\pi$. Remember that the propagator is given by 

\begin{align*}
U_{\mathrm{sw}}(\tau) &= \eup^{-\iup\tau\sum\limits_{i}s^{z}_{i}s^{z}_{i+1}}\eup^{-\iup\tau \sum\limits_{i}s^{z}_{i}} \eup^{-\iup\tau\sum\limits_{i}s^{x}_{i}}\\
&=\left(\prod\limits_{i}\eup^{-\iup\tau s^{z}_{i}s^{z}_{i+1}}\right)\left(\prod\limits_{i}\eup^{-\iup\tau s^{z}_{i}}\right)\left(\prod\limits_{i}\eup^{-\iup\tau s^{x}_{i}}\right)
\end{align*}

The single particle terms can be evaluated using the rotation formula 

$$ \prod\limits_{i}\eup^{-\iup\tau s^{x/z}_{i}} = \prod\limits_{i}\left(\cos(\frac{\tau}{2})I-\iup\sin(\frac{\tau}{2})\sigma^{x/z}_{i}\right) $$

and two particle terms by writing out the matrix in the computational basis

$$ \eup^{-\iup\tau\sum\limits_{i}s^{z}_{i}s^{z}_{i+1}} =\prod\limits_{i}\begin{pmatrix}
\eup^{-\iup\frac{\tau}{4}} & 0 & 0 & 0 \\ 
0 & \eup^{\iup\frac{\tau}{4}} & 0 & 0 \\ 
0 & 0 & \eup^{\iup\frac{\tau}{4}} & 0 \\ 
0 & 0 & 0 & \eup^{-\iup\frac{\tau}{4}}
\end{pmatrix}_{i, i+1}, $$

where $i, i + 1$ denote the spins upon which the matrix acts (note that at the boundary $i=L$ this has to be understood rather formally). Using this one can evaluate the propagator at multiples of $2\pi$ easily

\begin{itemize}
\item $U_{\mathrm{sw}}(8\pi)=I$ \small{(thus the entire angle diagram repeats between $ 8n\pi $ intervals)}
\item $U_{\mathrm{sw}}(4\pi + 8n\pi)=(-1)^{L}I.$ 
\item $U_{\mathrm{sw}}(2\pi + 8n\pi)=\eup^{-2\pi\iup\sum\limits_{i}s^{z}_{i}s^{z}_{i+1}}$ \small{$\Rightarrow \Heff(2\pi) \sim \sum\limits_{i}s^{z}_{i}s^{z}_{i+1}.$}
 \end{itemize}
 
The situation at $\pi$ is more difficult, since the single particle terms do not vanish but instead combine to $ \iup^{L}\prod\limits_{i}\sigma^{y}_{i}$. This means that the full propagator is given by
$$U_{\mathrm{sw}}(\pi)=\iup^{L} \eup^{-\iup\tau\sum\limits_{i}s^{z}_{i}s^{z}_{i+1}}\prod\limits_{i}\sigma^{y}_{i}.$$
We are unable to find a full expression for the Floquet Hamiltonian, but can prove that this propagator has a period 2 dynamics for even system sizes, meaning that $U_{\mathrm{sw}}^{2}(\pi)=I$, thus it is a toy example of a (stable but fine-tuned) discrete time crystal. For this let us work in the computational basis (product states along z-axis): $\prod\limits_{i}\sigma^{y}_{i}$ is completely anti-diagonal and the other term completely diagonal. One can readily verify that $ U^{2} _{\mathrm{sw}}$ is then a diagonal matrix with entries
\begin{align*}
\left( U^{2}_{\mathrm{sw}}(\pi) \right)_{i, i} &=  \left( \prod\limits_{i}\sigma^{y}_{i} \right)_{i, i'} \left(\prod\limits_{i}\sigma^{y}_{i} \right)_{i', i}\\
& \times \left( \eup^{-i\tau\sum\limits_{i}s^{z}_{i}s^{z}_{i+1}} \right)_{i, i}  \left( \eup^{-i\tau\sum\limits_{i}s^{z}_{i}s^{z}_{i+1}} \right)_{i', i'},
\end{align*} 
 where $i'$ is the "complement" i.e. the Hilbert space dimension minus $i$ which is also the state with all spins flipped. Due to the properties of $\sigma^{y}$ under spin flips the product of corresponding terms gives the identity. The interaction term is invariant under spin flips, therefore the entire expression is given by
 $$ \left( \eup^{-i2\tau\sum\limits_{i}s^{z}_{i}s^{z}_{i+1}} \right)_{i, i} =  \eup^{-i\frac{\tau}{2}\left(\sum\limits_{i}\sigma^{z}_{i}\sigma^{z}_{i+1}\right)_{i, i}} $$
 The matrix elements of $ \sum\limits_{i}\sigma^{z}_{i}\sigma^{z}_{i+1} $ are $ -2\ell +2k $, with $L=2\ell$ and $k$ being the number of kinks on top of the fully polarized state. Substituting this and some straightforward algebra leads to the claimed result that $U^{2}_{\mathrm{sw}}(\pi)=I$. On a final note we would like to point out that the Floquet Hamiltonian at $ \pi $ is not a simple single particle operator, since for some product states we observe an increase in the bipartite entanglement entropy upon action with the propagator.

\section{\label{app: hhz freq} Floquet Hamiltonian with Frequency Dependent Couplings}

We want to verify that the lowest order term in the Floquet Hamiltonian, in the protocol with frequency dependent couplings, is significantly different from the average Hamiltonian $H_{0}=\frac{1}{2}(X+Z+ZZ)$. As discussed in Sec.~\ref{sec: freq dep} an analytic formula for the lowest order term is not available, therefore we construct an approximation numerically. We compute the full propagator and then construct the Floquet Hamiltonian via full diagonalization. For sufficiently small half-periods this should largely coincide with the first term in the high-frequency expansion. In principle one can now obtain the coefficient of any operator (for example $X$) by using an appropriate scalar product. However, here we use the energy density $ \epsilon $ of the product states along the x-, y-, and z-axis to compare the contributions of operators consisting solely of $X$, $Y$ or $Z$ terms, since for these states the expectation value of all mixed terms vanishes. We find consistent results for $\tau \lessapprox 0.01 $ and system sizes $L=12-18$, which are summarized in Tab.~\ref{tab: energy density}.

\begin{table}[htbp]
\begin{tabular}{c|c|c|c|c|c|c|}
State & $\ket{z, +}$ & $\ket{z, -}$ & $\ket{x, +}$ & $\ket{x, -}$ & $\ket{y, +}$ & $\ket{y, -}$ \\ 
\hline 
$\epsilon_{0}$ & 3/8 & -1/8 & 1/4 & -1/4 & 0 & 0 \\ 
\hline 
$\epsilon$ & 0.258 & -0.069 & 0.184 & 0.084 & 0.326 & -0.333 \\ 
\hline 
\end{tabular} 
\caption{\label{tab: energy density} Energy density of product states with respect to the average Hamiltonian ($\epsilon_{0}$) and the full lowest order term $\epsilon$. The results are based on $L=18$ and $ \tau=10^{-3}$.}
\end{table}

 As one can see the energy densities of the products states along the x- and z-axis change significantly and in the case of x-states do not have opposite signs anymore, hence there must be additional operators with even number of $X$ terms. The most striking change however is seen from the y-states, which go from a vanishing energy density to the largest/smallest one. Since they are almost the negative of another, the largest contribution comes from operators with odd number of $Y$. Finally, we also observe that the y-states are almost at the very edge of the spectrum, therefore we conclude that the Floquet Hamiltonian is significantly different from the average and that the heating rates, although looking similar are the result of a truly different dynamics. 

\section{\label{app: ye} Spin Chain with Spin Flip Symmetry Breaking}

Additionally, to the driven spin chain in the main part, we consider a model of a spin chain, which has been studied in the context of heating in Floquet systems \cite{machado_exponentially_2019, ye_emergent_2020} and also can be regarded as an example system with weak symmetry breaking. The average and drive are given by
$$H_{0} =h_{x}X+J_{z}ZZ+J_{x}XX, \quad V = h_{y}Y+h_{z}Z,$$
with coefficients $h_{x}=0.42, h_{y}=0.34, h_{z}=0.26, J_{x}=3, J_{z}=4$, where the letters denote spin one-half operators analogous to the main text. Different from the cited works, we use periodic boundary conditions, which however does not seem to change the observed heating rates as well as further conclusions in this section. The average Hamiltonian is then invariant under translations, spatial reflection and spin flips about the z-axis generated by the flip operator 
$$ F=\prod\limits_{i=1}^{L}\sigma^{x}_{i}.$$
The drive has the same spatial symmetries but is not invariant under spin flips, which leads to a (weak) breaking of this symmetry, whose effects we will observe in the spectroscopic approach. Due to the spatial symmetries we can again focus on the zero momentum and positive parity sector. 

\begin{figure*}[t]
\subfloat{\label{fig: ye colored}\includegraphics[scale=1]{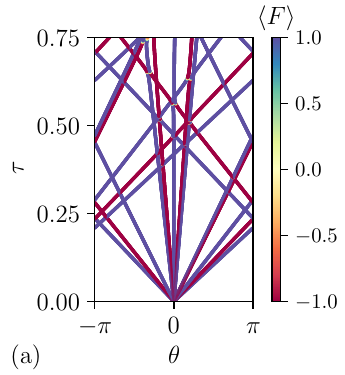}}
\hfill
\subfloat{\label{fig: ye transfer}\includegraphics[scale=1]{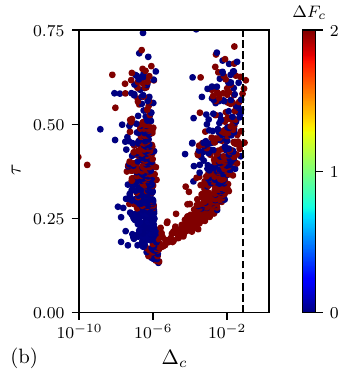}}
\hfill
\subfloat{\label{fig: ye heating} \includegraphics[scale=1]{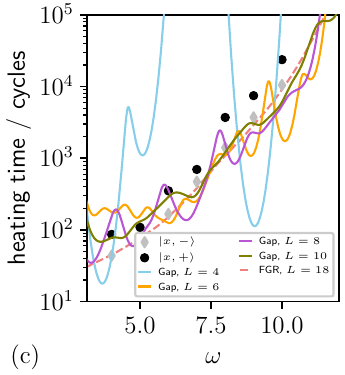}}
\caption{\label{fig: ye panel} Avoided crossing spectroscopy of spin chain with weak symmetry breaking: (a) Eigenangles with $L=6$ colored by spin flip parity. The states are close to fixed parity states, but crossings between both sectors are possible. (b) Gaps colored by spin flip transfer with $L=10$. For high-frequency only gaps between different sectors have an appreciable magnitude and contribute to heating. With increasing $\tau$ the ordering is changed (see discussion) and it appears that the transfer is independent of the sectors. (c) Heating times based on real-time simulations with $L=24$ (symbols) and the FGR as well as the gap data evaluated with $dE=0.5$ (lines).}
\end{figure*}

Anticipating a near-conservation of spin flip parity ($\braket{F}$) we color the levels in Fig.~\ref{fig: ye panel} by $\braket{F}$ instead of $\epsilon_{0}$ and can observe clearly in Fig.~\ref{fig: ye colored} that the eigenstates are (almost) fixed parity states and in Fig.~\ref{fig: ye transfer} that the matrix elements of states with the same parity are significantly smaller than the ones between different parity (this result can be explained by the oddness of the drive under spin flips \cite{ye_emergent_2020}). In fact, the small gaps seem to be limited by resolution and might be much smaller in actuality, however they will not affect the heating rate in any case. Looking closely at the Floquet diagrams, it seems that the smaller gaps should be attributed to same parity states for all $\tau$ values and not be mixed as in the figure. The observed mixing should rather be understood as illustrating the effects of ghost gaps, as discussed in the main part. 

Furthermore, upon inspection of the diagrams, we observe phenomena that we call "nested crossings": here there is a switching between two states which are not neighbors in the Floquet diagram, but rather separated by a middle state which seems unaffected. We have not observed such crossings in the driven chain and suspect that they are related to the (nearly) spin flip symmetry. Anyway, this seems to not affect the heating times displayed in Fig.~\ref{fig: ye heating} and further supports the robustness of the method, while also suggesting the study of crossings with near symmetries as a potential future direction.

\bibliography{bib_crossings}
\end{document}